\documentclass[12pt]{article}
\usepackage{amsfonts,amssymb,amsmath}
\usepackage{graphicx}
\usepackage[super,sort,compress]{cite} 

\usepackage{chemfig}
\usepackage{color}
\usepackage{array}

\def\be{\begin{align}}
\def\ds{\displaystyle}

\begin{document}

\title{Telling apart \textsl{Felidae} and \textsl{Ursidae} from the distribution of nucleotides in mitochondrial DNA
}

\author{Andrij Rovenchak\\
Department for Theoretical Physics,\\
Ivan Franko National University of Lviv, \\
12 Drahomanov St., Lviv, UA-79005, Ukraine
\\
andrij.rovenchak@gmail.com}

\maketitle

\begin{abstract}
Rank--frequency distributions of nucleotide sequences in mitochondrial DNA are defined in a way analogous to the linguistic approach, with the highest-frequent nucleobase serving as a whitespace. For such sequences, entropy and mean length are calculated. These parameters are shown to discriminate the species of the \textit{Felidae} (cats) and \textit{Ursidae} (bears) families. From purely numerical values we are able to see in particular that giant pandas are bears while koalas are not. The observed linear relation between the parameters is explained using a simple probabilistic model. The approach based on the nonadditive generalization of the Bose-distribution is used to analyze the frequency spectra of the nucleotide sequences. In this case, the separation of families is not very sharp. Nevertheless, the distributions for \textit{Felidae} have on average longer tails comparing to \textit{Ursidae}


\end{abstract}

\section{Introduction}\label{sec:Intro}

Approaches of statistical physics proved to be efficient tools for studies of systems of different nature containing many interacting agents.
Applications cover a vast variety of subjects, from voting models \cite{Rozanova&Boguna:2017,Pickering&Lim:2017}, language dynamics \cite{Burridge:2017,Lipowska&Lipowski:2017}, and wealth distribution \cite{Benisty:2017} to dynamics of infection spreading \cite{Ottino-Loffler_etal:2017} and cellular growth \cite{DeMartino_etal:2017}.

Studies of deoxyribonucleic acid (DNA) and genomes are of particular interest as they can bridge several scientific domains, namely, biology, physics, and linguistics \cite{Furusawa&Kaneko:2003,Li_etal:2009,Harmston_etal:2010,Eroglu:2014,Eroglu:2015}. Such an interdisciplinary nature of the problem might require a brief introductory information as provided below.

DNA molecule is composed of two polynucleotide chains (called stands) containing four types of nucleotide bases: adenine (A), cytosine (C), guanine (G), and thymine (T) \cite[p.~175]{Alberts_etal:2015}. The bases attached to a phosphate group form nucleotides. Nucleotides are linked by covalent bonds within a strand while there are hydrogen bonds between strands (A links T and C links with G) forming base pairs (bp). Typically, DNA molecules count millions to hundred millions base pairs.

Mitochondrial DNA (mtDNA) is a closed-circular, double-stranded molecule containing, in the case of mammals, 16--17 thousand base pairs \cite{Taanman:1999}. It is thought to have bacterial evolutionary origin, so mtDNA might be considered as a nearly universal tool to study all eukaryotes. Mitochondrial DNA of animals is mostly inherited matrilineally and encodes the same gene content \cite{Wolstenholme:1992}. Due to quite short mtDNA sizes of different organisms, it is not easy to collect reliable statistical data for typical nucleotide sequences, like genes or even codons containing three bases.

Power-law distributions characterize rank--frequency relations of various units. Originally observed by Estoup in French \cite{Estoup} and Condon in English \cite{Condon:1928} and better known from the works of Zipf \cite{Zipf:1935,Zipf:1949}, such a dependence is observed not only in linguistics \cite{Ferrer_i_Cancho&Sole:2001,Montemurro:2001,Ha_etal:2002,Piantadosi:2014,Williams_etal:2015} but for complex systems in general \cite{Gerlach_etal:2016,Holovatch_etal:2017} as manifested in urban development \cite{Ghosh_etal:2014,Arcaute_etal:2014,Jiang_etal:2015}, income distribution \cite{Benisty:2017,Aoki&Nirei:2017}, genome studies \cite{Furusawa&Kaneko:2003,Ogasawara,Sheinman_etal:2016}, and other domains \cite{Zanette:2006,Aitchison_etal:2016}.

The aim of this paper is to propose a simple approach based on frequency studies of nucleotide sequences in mitochondrial DNA, which would make it possible to define a set of parameters separating biological families and genera. The analysis is made for two carnivoran families of mammals, \textit{Felidae} (including cats) \cite{OBrien_etal:2008} and \textit{Ursidae} (bears) \cite{Yu_etal:2007}. Applying the proposed approach, we are able to discriminate the two families and also draw conclusions whether some other species belong to these families or not.

The paper is organized as follows. In Section~\ref{sec:Sequences}, the nucleotide sequences used in further analysis are defined. Section~\ref{sec:Params} contains the analysis based on rank--frequency distributions of these nucleotide sequences. In Section~\ref{sec:Model}, a simple model is suggested to describe the observed frequency data and relations between the respective parameters. Frequency spectra corresponding to rank--frequency distributions are analyzed in Section~\ref{sec:Spectra}. Conclusions are given in Section~\ref{sec:Concl}.

\section{Nucleotide sequences}\label{sec:Sequences}
The four nucleobases forming mtDNA are the following compounds: 

\bigskip\noindent
Adenine (C$_5$H$_5$N$_5$):\quad
{\tiny
\chemfig{[:-36]*5(-N(-H)-*6(-N=-N=(-NH_2)--)--N=)}
}

\bigskip\noindent
Cytosine (C$_4$H$_5$N$_3$O):\quad
{\tiny
\chemfig{H-[:30]N*6(-(=O)-N=(-NH_2)-=-)}
}

\bigskip\noindent
Thymine (C$_5$H$_6$N$_2$O$_2$):\quad
{\tiny
\chemfig{H-[:30]N*6(-(=O)-N(-H)-(=O)-(-CH_3)=-)}
}

\bigskip\noindent
Guanine (C$_5$H$_5$N$_5$O):\quad
{\tiny
\chemfig{[:-36]*5(-N(-H)-*6(-N=(-NH_2)-N(-H)-(=O)--)--N=)}
}

\bigskip
In Table~\ref{tab:Nucleobases}, the information is given about the absolute and relative frequency of each nucleotide in mtDNA of the species analyzed in this work. The data are collected from databases of The National Center for Biotechnology Information \cite{NCBIH:www}.

\begin{table}[h]
\caption{Distribution of nucleobases in species}\label{tab:Nucleobases}

{\scriptsize
\begin{tabular}{>{\it }l >{(}l<{)} c c c c c c c c c}
\hline
\multicolumn{2}{c}{Name}	&	Size,\rule{0pt}{1.2em}	
&	\multicolumn{2}{c}{A}	
&	\multicolumn{2}{c}{C}	
&	\multicolumn{2}{c}{T}
&	\multicolumn{2}{c}{G}	\\
\cline{1-2}
\cline{4-5}
\cline{6-7}
\cline{8-9}
\cline{10-11}
{\rm Latin}	&	common	&	bp	&	abs.	&	rel.	&	abs.	&	rel.	&	abs.	&	rel.	&	abs.	&	rel.	\\
\hline
\textbf{Felidae:}\rule{0pt}{1.2em}\\[3pt]
Acinonyx jubatus	&	cheetah	&	17047	&	5642	&	0.331	&	4397	&	0.258	&	4693	&	0.275	&	2315	&	0.136	\\
Felis catus	&	domestic cat	&	17009	&	5543	&	0.326	&	4454	&	0.262	&	4606	&	0.271	&	2406	&	0.141	\\
Leopardus pardalis	&	ocelot	&	16692	&	5493	&	0.329	&	4454	&	0.267	&	4457	&	0.267	&	2288	&	0.137	\\
Lynx lynx	&	European lynx	&	17046	&	5509	&	0.323	&	4615	&	0.271	&	4488	&	0.263	&	2434	&	0.143	\\
Panthera leo	&	lion	&	17054	&	5448	&	0.319	&	4520	&	0.265	&	4608	&	0.270	&	2478	&	0.145	\\
Panthera onca	&	jaguar	&	17049	&	5447	&	0.319	&	4509	&	0.264	&	4627	&	0.271	&	2466	&	0.145	\\
Panthera pardus	&	leopard	&	16964	&	5397	&	0.318	&	4508	&	0.266	&	4592	&	0.271	&	2467	&	0.145	\\
Panthera tigris	&	tiger	&	16990	&	5418	&	0.319	&	4513	&	0.266	&	4581	&	0.270	&	2478	&	0.146	\\
Puma concolor	&	puma	&	17153	&	5643	&	0.329	&	4456	&	0.260	&	4685	&	0.273	&	2369	&	0.138	\\
\hline	
\\
\textbf{Ursidae:}\\[3pt]
Ailuropoda melanoleuca	&	giant panda	&	16805	&	5338	&	0.318	&	4000	&	0.238	&	4949	&	0.294	&	2518	&	0.150	\\
Helarctos malayanus	&	sun bear	&	16783	&	5232	&	0.312	&	4295	&	0.256	&	4678	&	0.279	&	2578	&	0.154	\\
Melursus ursinus	&	sloth bear	&	16817	&	5184	&	0.308	&	4364	&	0.259	&	4619	&	0.275	&	2650	&	0.158	\\
Tremarctos ornatus	&	spectacled bear	&	16762	&	5246	&	0.313	&	4348	&	0.259	&	4583	&	0.273	&	2585	&	0.154	\\
Ursus americanus	&	Am. black bear	&	16841	&	5244	&	0.311	&	4223	&	0.251	&	4760	&	0.283	&	2614	&	0.155	\\
Ursus arctos	&	brown bear	&	17020	&	5258	&	0.309	&	4355	&	0.256	&	4731	&	0.278	&	2676	&	0.157	\\
Ursus maritimus	&	polar bear	&	17017	&	5253	&	0.309	&	4346	&	0.255	&	4726	&	0.278	&	2692	&	0.158	\\
Ursus spelaeus	&	cave bear	&	16780	&	5272	&	0.314	&	4256	&	0.254	&	4703	&	0.280	&	2549	&	0.152	\\
Ursus thibetanus	&	Asian black bear	&	16794	&	5234	&	0.312	&	4284	&	0.255	&	4683	&	0.279	&	2593	&	0.154	\\
\hline																		\\
{\bf Other:}\\[3pt]
Ailurus fulgens	&	red panda	&	16493	&	5426	&	0.329	&	4001	&	0.243	&	4863	&	0.295	&	2203	&	0.134	\\[6pt]
Phascolarctos cinereus	&	koala	&	16357	&	5854	&	0.358	&	4144	&	0.253	&	4501	&	0.275	&	1858	&	0.114	\\[6pt]
Drosophila melanogaster	&	drosophila	&	19517	&	8152	&	0.418	&	2003	&	0.103	&	7883	&	0.404	&	1479	&	0.076	\\
\hline
\end{tabular}
}
\end{table}

Note the highest frequency of adenine in mtDNA for all species from Table~\ref{tab:Nucleobases}. There is a well-known linguistic analogy for between distribution of DNA units \cite{Searls:1992,Ji:1999}. The ``alphabet'' can be defined as a set of four nucleobases or 24 aminoacids. In the present paper, we will not advance to further levels of ``linguistic'' organization of DNA \cite{Ji:1999} but introduce a new one in a way similar to \cite{Rovenchak&Buk:2018}. This is done in the following fashion: as adenine is the most frequent nucleobase, one can treat it as a whitespace separating sequences of other nucleobases (C, T, and G). While a special role of adenine might be attributed to oxygen missing in its structure formula (given at the beginning of this Section), neither this nor other arguments dealing with specific nucleobase properties will be considered in the proposed approach, which is aimed to be a simple frequency-based analysis. For convenience, an empty element (between two As) will be denoted as `X'. So, the sequence from the mtDNA of \textit{Ailuropoda melanoleuca} (giant panda)

\centerline{{\tt ATACTATAAATCCACCTCTCATTTTATTCACTTCATACATGCTATTACAC}}

\noindent translates to

\centerline{{\tt X T CT T X X TCC CCTCTC TTTT TTC CTTC T C TGCT TT C C}}

The obtained ``words'' (sequences between spaces) are the units for the analysis in the present work.

\section{Analysis of rank--frequency dependences}\label{sec:Params}
The rank--frequency dependence is compiled for the defined nucleobase sequences in a standard way, so that the most frequent unit has rank 1, the second most frequent unit has rank 2 and so on. Units with equal frequencies are arbitrarily ordered within a consecutive range of ranks. Samples are shown in Table~\ref{tab:rf}.

\begin{table}
\caption{Rank--frequency list for some species}\label{tab:rf}
\begin{tabular}{r| l r| l r| l r| l r}
\hline
rank&\multicolumn{2}{c|}{\textit{Felis catus}}
&\multicolumn{2}{c|}{\textit{Panthera leo}}
&\multicolumn{2}{c|}{\textit{A. melanoleuca}}
&\multicolumn{2}{c}{\textit{Ursus arctos}}\\
$r$	&	seq.	&	$f_r$	&	seq.	&	$f_r$	&	seq.	&	$f_r$	&	seq.	&	$f_r$	\\
\hline
1	&	X	&	1725	&	X	&	1691	&	X	&	1241	&	X	&	1586	\\
2	&	C	&	483	&	T	&	484	&	T	&	483	&	T	&	410	\\
3	&	T	&	478	&	C	&	449	&	C	&	365	&	C	&	375	\\
4	&	G	&	216	&	G	&	200	&	G	&	220	&	G	&	214	\\
5	&	CT	&	179	&	CT	&	170	&	CT	&	186	&	CT	&	165	\\
6	&	TT	&	163	&	TT	&	145	&	TT	&	178	&	TC	&	138	\\
7	&	TC	&	143	&	TC	&	137	&	TC	&	118	&	TT	&	134	\\
8	&	CC	&	116	&	CC	&	117	&	TG	&	99	&	TG	&	101	\\
9	&	TG	&	99	&	TG	&	93	&	GC	&	92	&	CC	&	92	\\
10	&	GC	&	81	&	GC	&	89	&	GT	&	84	&	GT	&	86	\\
11	&	GG	&	79	&	GG	&	81	&	GG	&	83	&	GC	&	85	\\
12	&	GT	&	78	&	GT	&	69	&	CC	&	82	&	GG	&	85	\\
13	&	CG	&	48	&	CGT	&	53	&	CG	&	44	&	CGC	&	67	\\
14	&	CCC	&	39	&	CG	&	45	&	CTT	&	44	&	CGTGT	&	53	\\
15	&	CCT	&	38	&	CCC	&	39	&	TTT	&	43	&	CG	&	45	\\
16	&	CGT	&	38	&	CCT	&	35	&	CCT	&	41	&	TTT	&	44	\\
17	&	GCC	&	37	&	GCC	&	34	&	GCT	&	35	&	CTT	&	35	\\
18	&	TTT	&	36	&	CTC	&	33	&	CGTGT	&	34	&	GCT	&	35	\\
19	&	TCT	&	32	&	TTC	&	33	&	TCC	&	32	&	TCC	&	33	\\
20	&	CTT	&	31	&	TTT	&	31	&	TCT	&	30	&	CTC	&	31	\\
21	&	CTC	&	30	&	GCT	&	30	&	TTC	&	30	&	TTC	&	29	\\
22	&	GCT	&	30	&	CTT	&	29	&	GCC	&	28	&	CCT	&	28	\\
23	&	TCC	&	26	&	TCT	&	27	&	CGC	&	26	&	TCT	&	28	\\
24	&	TCCT	&	24	&	TGT	&	23	&	CTC	&	25	&	GCC	&	26	\\
25	&	TGT	&	23	&	TCC	&	22	&	GTT	&	24	&	CCC	&	25	\\
\hline
\end{tabular}
\end{table}

\begin{figure}[h]
\centerline{\includegraphics[scale=0.8]{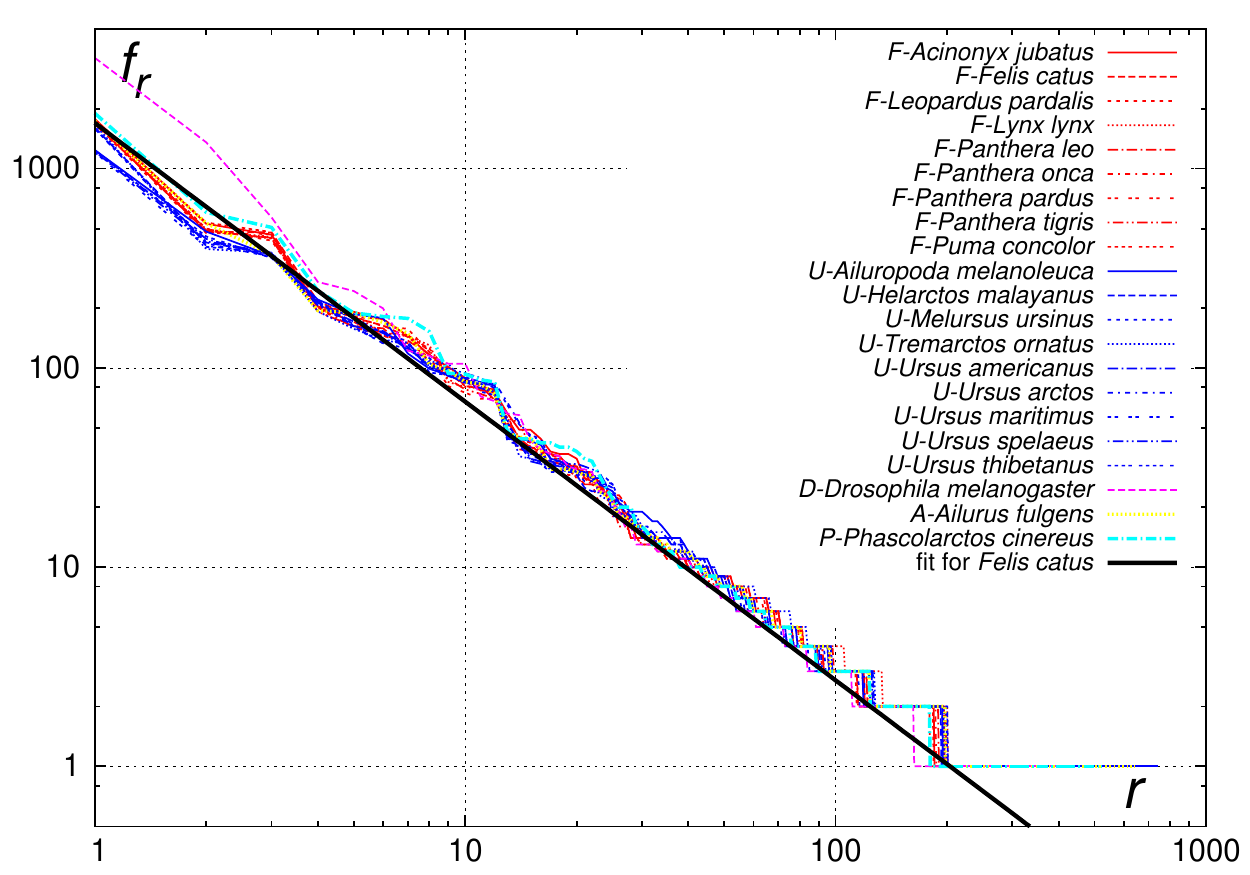}}
\caption{Rank--frequency dependence.
}\label{fig:rank}
\end{figure}

The rank--frequency dependences follow Zipf's law very precisely (see Fig.~\ref{fig:rank}), so they can be modeled by
\be
f_r = \frac{C}{r^\alpha}.
\end{align}

\noindent
The normalization condition
\be
\sum_r f_r =\sum_{r=1}^\infty \frac{C}{r^\alpha} = N
\end{align}
yields
\be
C=\frac{N}{\zeta(\alpha)},
\end{align}
where $\zeta(\alpha)$ is Riemann's zeta-function.

Entropy $S$ can be defined in a standard way,
\be
S = -\sum_r p_r \ln p_r,
\end{align}
where relative frequency $p_r = f_r/N$ and the summation runs over all the ranks.

After simple manipulations we obtain the following expression for entropy:
\be\label{eq:S0}
S = \ln\zeta(\alpha) - \alpha \frac{\zeta'(\alpha)}{\zeta(\alpha)},
\end{align}
where the sum
\be
\sum_{r=1}^\infty \frac{\ln r}{r^\alpha} = -\zeta'(\alpha).
\end{align}
Due to a weak convergence ($1<\alpha<2$) it might be reasonable to consider finite summations over $r$ and hence the incomplete zeta-functions
\be
\sum_{r=1}^R \frac{1}{r^\alpha} &= \zeta(\alpha)-\zeta(\alpha,R+1),\\
\sum_{r=1}^R \frac{\ln r}{r^\alpha} &= -\zeta'(\alpha)+\zeta_\alpha(\alpha,R+1),
\end{align}
where
\be
\zeta_\alpha(\alpha,R+1) \equiv \frac{\partial \zeta(\alpha,R+1)}{\partial \alpha}.
\end{align}
In this case, the entropy equals
\be\label{eq:S}
S = \ln[\zeta(\alpha)-\zeta(\alpha,R+1)] - 
\alpha \frac{\zeta'(\alpha)-\zeta_\alpha(\alpha,R+1)}{\zeta(\alpha)-\zeta(\alpha,R+1)}.
\end{align}

\begin{figure}[h]
\centerline{\includegraphics[scale=0.8]{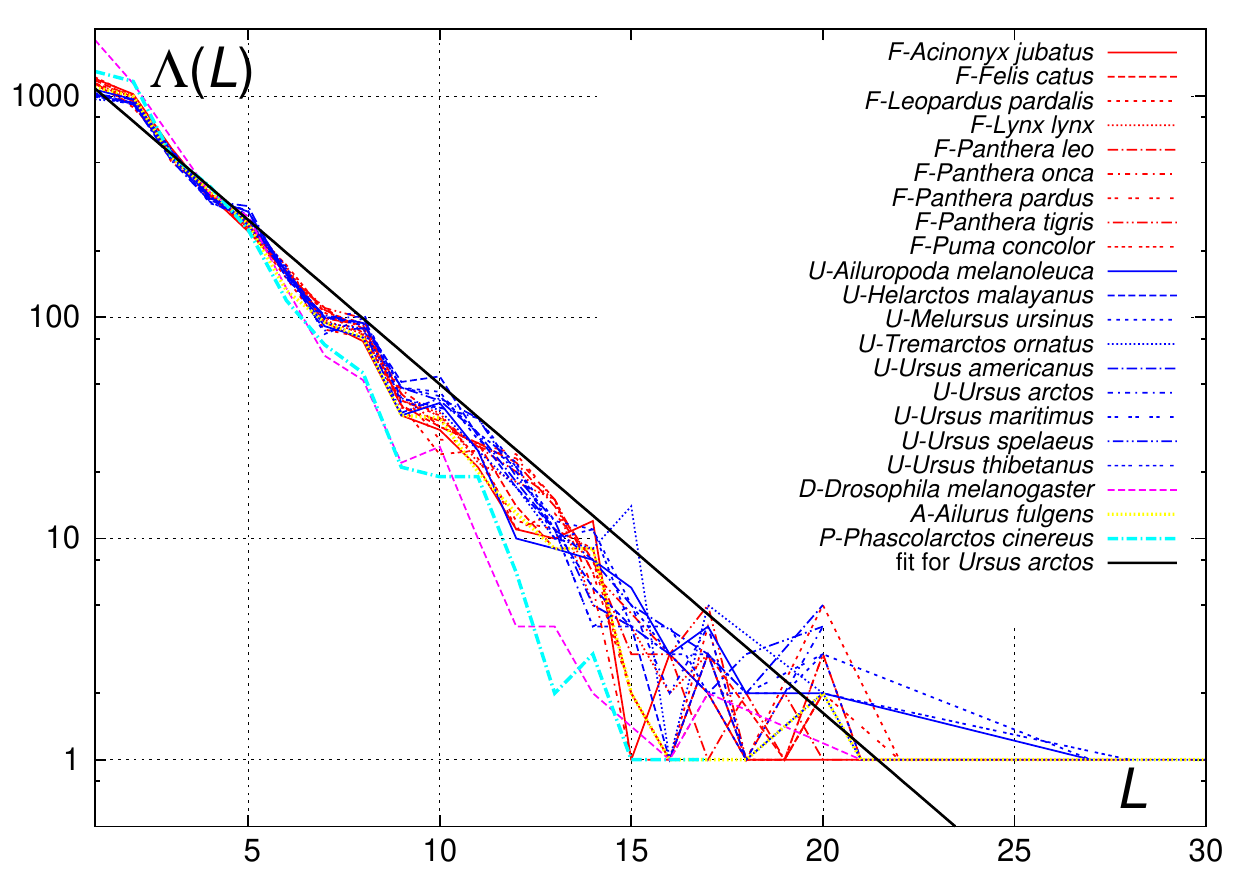}}
\caption{Sequence lengths. The first capital letter before the hyphen serves to distinguish families.
}\label{fig:length}
\end{figure}

Mean sequence length
\be
\langle L\rangle = \frac{1}{N}\sum_L L \Lambda(L).
\end{align}
The length distribution $\Lambda(L)$ can be approximately treated as a linear dependence in log-linear plot (see Fig.~\ref{fig:length}), so
\be
\ln \Lambda(L) = -KL+B.
\end{align}
Applying the normalization condition
\be
\sum_{L=0}^\infty \Lambda(L) = N 
\end{align}
we obtain
\be\label{eq:Lambda}
\Lambda(L) = N\left(1-e^{-K}\right)e^{-KL}
\end{align}
and
\be\label{eq:L}
\langle L\rangle = \frac{1}{e^K-1}\simeq \frac{1}{K},
\end{align}
the latter approximation corresponding to $K\ll1$.

\begin{figure}[h]
\centerline{\includegraphics[scale=0.8]{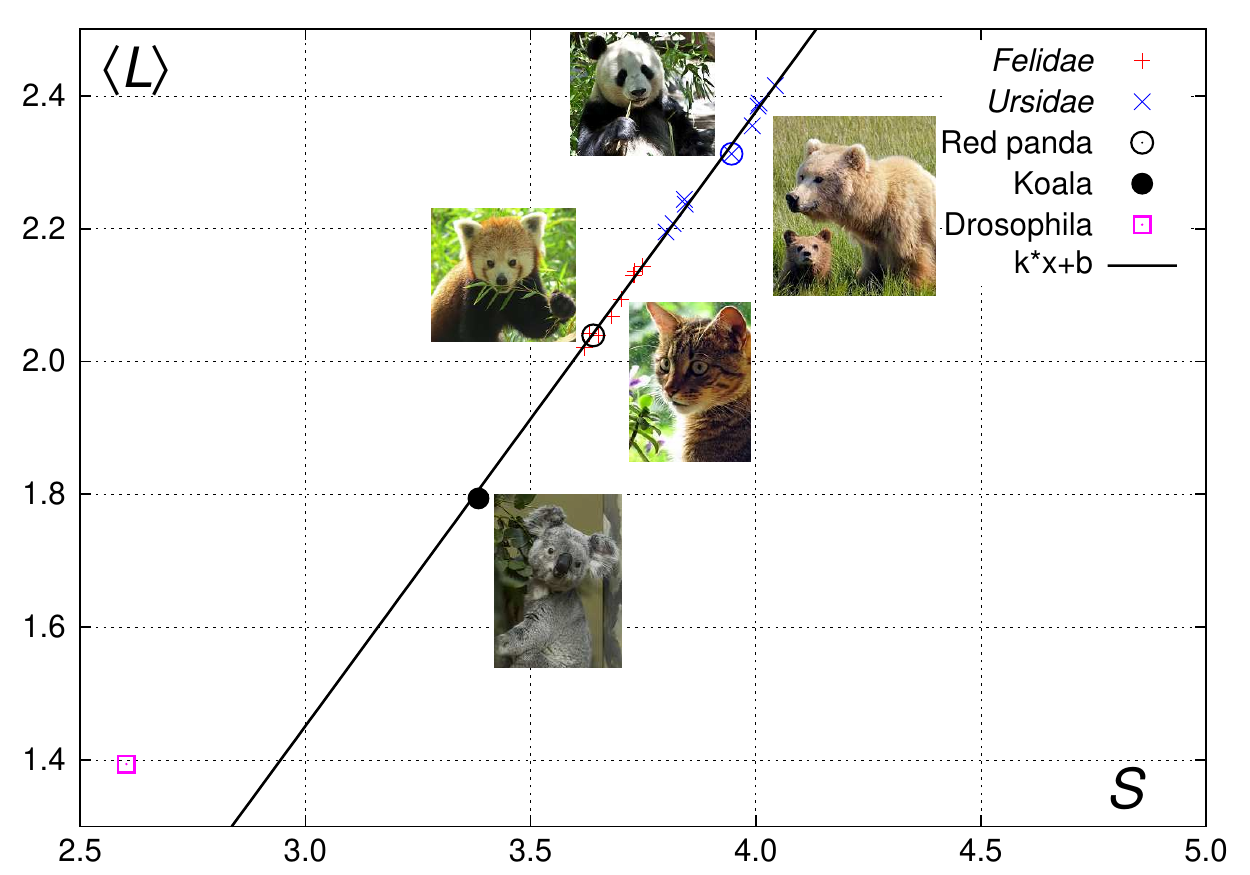}}
\caption{Entropy $S$ and mean length $\langle L\rangle$ for families and species analyzed in the present work.
}\label{fig:S_x}
\end{figure}

As shown in Fig.~\ref{fig:S_x}, the values of $S$ and $\langle L\rangle$ concentrate along a straight line,
\be
\langle L\rangle = k S + b,\quad
\textrm{with}\quad k=0.924\pm 0.015,\ \ b=-1.32\pm0.06.
\end{align}

Images in Fig.~\ref{fig:S_x} serve to illustrate partial results dealing with some popular misbeliefs about bears. First of all, one clearly sees a large distance between bears and koala. The latter is known also as \textit{koala bear} \cite{Leitner&Sieloff:1998} or \textit{marsupial bear} in many languages; the genus name (\textit{Phascolarctos}) itself is composed of Greek words $\varphi\acute{\alpha}\sigma\kappa\omega\lambda o\varsigma$ `leathern bag' and \raisebox{0.2ex}{\footnotesize\textrm{'}}$\!\!\acute{\alpha}\rho\kappa\tau o\varsigma$ `bear' \cite[p.~529]{Palmer:1904}. Despite the name and appearance, koalas are clearly not bears. On the other hand, we are able to confirm that giant pandas are bears and red pandas belong to a different genus. It would be incorrect however to place red pandas within \textit{Felidae} based solely on the parameter values. Out of sheer curiosity, note some local names for pandas in Nepali
\raisebox{-0.6ex}{\includegraphics[scale=0.7]{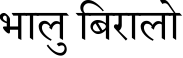}}
\textit{bh\=alu bir\=al\=o} 
and Chinese
\raisebox{-0.2ex}{\includegraphics[scale=0.7]{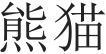}} 
\textit{xi\'ongm\=ao}
meaning `bear-cat', cf. \cite[p.~143]{Acharya:1986} and \cite[p.~12]{Glatston:2011}.

\begin{figure}[h]
\centerline{\includegraphics[scale=0.8]{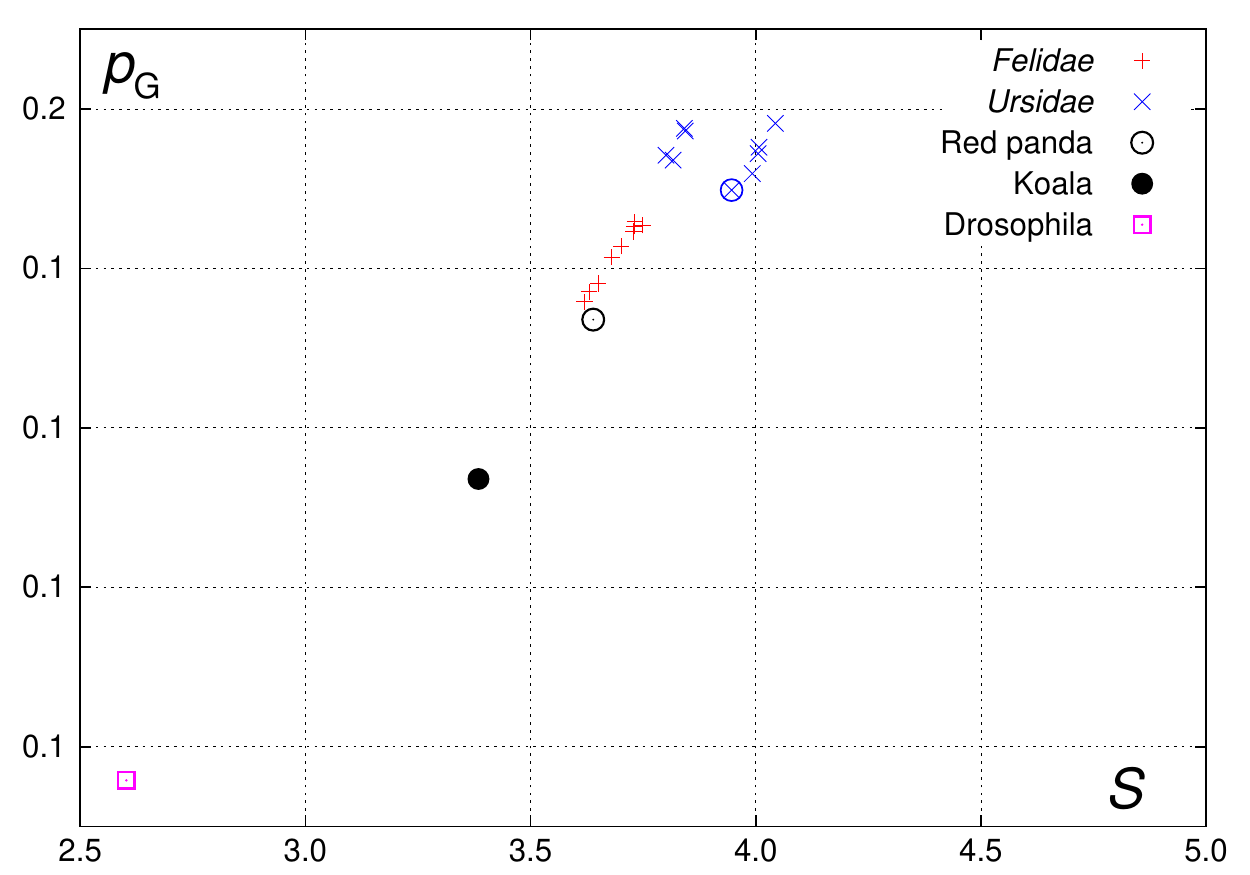}}
\caption{Entropy $S$ and relative frequency of guanine $p_{\rm G}$ for families and species analyzed in the present work.
}\label{fig:S_G}
\end{figure}

Another pair of parameters to distinguish the \textit{Felidae} and \textit{Ursidae} families can be chosen from Table~\ref{tab:Nucleobases}. It is clearly seen that the relative frequency of guanine $p_{\rm G}$ is a good discriminating parameter. The pairs of entropy $S$ and $p_{\rm G}$ are plotted in Fig.~\ref{fig:S_G}.

The data for a model organism in biological studies, \textit{Drosophila melanogaster} or common fruit fly, are shown for comparison and future references. We can observe that the parameters for this species differ significantly from those of the analyzed mammals. It can be considered as another confirmation that the proposed parameters can serve to distinguish families and genera.

\section{Random model}\label{sec:Model}
Assuming that the chain of nucleotides forming the mitochondrial DNA is long enough, one can propose the following simplified model for the distribution of the defined nucleotide sequences. As seen from Table~\ref{tab:Nucleobases}, relative frequencies of cytosine and thymine are nearly equal, the frequency of adenine is slightly larger, and the frequency of guanine is about twice smaller than that of cytosine or thymine. So, let the probability to find cytosine and thymine
\be
p_{\rm C}=p_{\rm T}=p,
\end{align}
for adenine, 
\be
p_{\rm A} = (1+a)p, \qquad a>0.
\end{align}
and for guanine
\be
p_{\rm G} = \frac12 p.
\end{align}
The normalization condition yields
\be
p=\frac{2}{7+2a}.
\end{align}
So, for a Markovian chain (i.e., a randomly generated sequence) we have the probabilities to find
\begin{tabbing}
an empty element X:\quad \= $p_{\rm A}p_{\rm A} = (1+a)^2p^2$\\[6pt]
a single nucleotide: \> 
$p_{\rm A}p_{\rm C}p_{\rm A} = 
p_{\rm A}p_{\rm T}p_{\rm A}= p_{\rm A}^2p,\quad
p_{\rm A}p_{\rm G}p_{\rm A}=\frac12 p_{\rm A}^2p$\\[6pt]
CC, CT, TC, TT: \> $p_{\rm A}^2p^2$\\[6pt]
CG, GC, TG, GC:\> $\frac12p_{\rm A}^2p^2$\\[6pt]
GG:\> $\frac14p_{\rm A}^2p^2$\\[6pt]
CCC, CCT, \ldots: \> $p_{\rm A}^2p^3$\\[6pt]
and so on.
\end{tabbing}

For simplicity, we will further give all the probabilities relative the the highest value $p_{\rm A}^2=(1+a)^2p^2$.

It is easy to show that the function $\Lambda(L)$ up to a constant factor $\Lambda_0$ equals

\begin{tabular}{ll}
$L$ & $\Lambda(L)$\\[0pt]
\hline
0 & $\Lambda_0\cdot1$\\[6pt]
1 & $\ds\Lambda_0\left(2p+\frac12p\right)$\\[12pt]
2 & $\ds\Lambda_0\left(4p^2+4\cdot\frac12p^2+\frac14p^2\right)$\\[12pt]
3 & $\ds\Lambda_0\left(8p^3+12\cdot\frac12p^2+6\cdot\frac14p^2+\frac18p^3\right)$\\[12pt]
\hline
\end{tabular}  

\bigskip\noindent
Generally,
\be
\Lambda(L) = 
\Lambda_0\sum_{\ell=0}^L\binom{L}{\ell}2^{L-\ell}\cdot\frac{p^L}{2^\ell}=
\Lambda_0\left(\frac52p\right)^L.
\end{align}
We thus obtain an exact exponential dependence as given by Eq.~(\ref{eq:Lambda}) with
\be
\Lambda(L) \propto e^{L\ln\frac52p}.
\end{align}
For $p=1/4$ this yields $K\simeq0.47$, i.e., $\Lambda(L) \propto e^{-0.47L}$.
The lowest value would correspond to $a=0$ so that $p_{\rm A} = p_{\rm C} = p = \frac27$ and $K\simeq0.34$.

The rank--frequency distribution corresponding to the proposed model would contain numerous plateaus at frequencies $p$, $\frac12p$, $p^2$, $\frac12p^2$, $\frac14p^2$, $p^3$, etc. Neglecting accidental degeneracies (which are possible, e.g., for $p=\frac14$), one can show that frequency $p^n$ corresponds to the range of ranks $(3^n-1)/2+1$ to $(3^n-1)/2+2^n$. Its midpoint $r=(3^n+2^n)/2$ thus corresponds to the absolute frequency $f_r={\rm const} \cdot p^n$. For $n$ large enough, $2r\simeq 3^n$ and
\be
f_r = {\rm const}\cdot r^{\frac{\ln p}{\ln3}}.
\end{align}
Depending on the values of $a$, this yields the Zipfian exponent $\alpha\simeq1.1$ to $1.3$, which is slightly lower than the observed scaling. 

\begin{figure}[h]
\centerline{\includegraphics[scale=0.80]{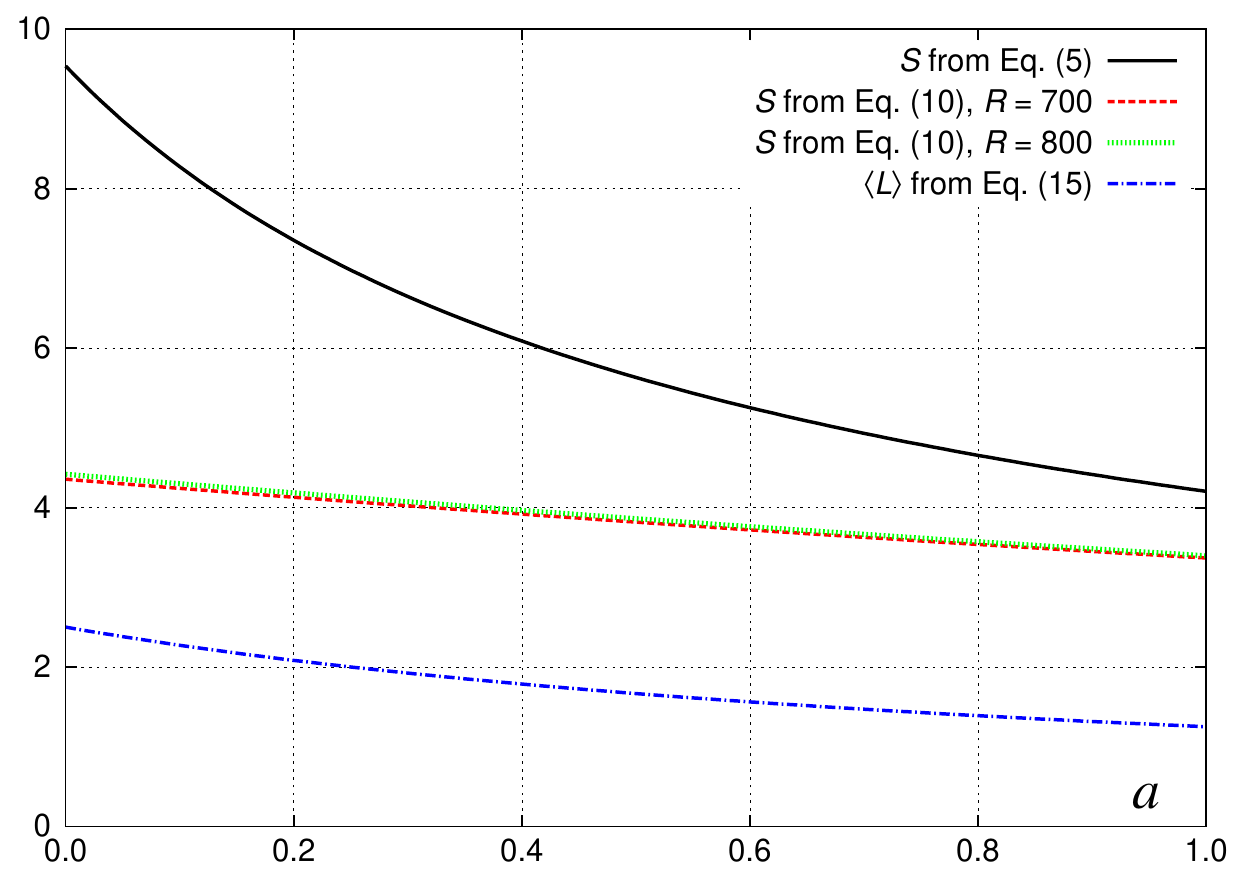}}
\caption{Entropy $S$ and mean sequence length $\langle L\rangle$ in the random model. Entropies for $R=700$ and $R=800$ nearly overlap.
}\label{fig:SLmodel}
\end{figure}

Entropy $S$ and mean length $\langle L\rangle$ calculated according to Eqs.~(\ref{eq:S}) and (\ref{eq:L}) are plotted in Fig.~\ref{fig:SLmodel}. Typical values of $R$ range from 689 from \textit{Ailuropoda melanoleuca} and \textit{Felis catus} to 741 for \textit{Melursus ursinus} and 742 for \textit{Panthera tigris}.

These quantities satisfy the following linear relation in the domain of $a\in[0;\frac12]$:
\begin{align*}
&\langle L\rangle \simeq 1.549\,S -4.245\qquad\textrm{for}\quad R=700,\\
&\langle L\rangle \simeq 1.496\,S -4.110\qquad\textrm{for}\quad R=800.
\end{align*}

It becomes thus clear that the proposed simple model can be used as the principal approximation requiring further adjustments to account for finer effects linked in particular with the exact relative numbers of nucleobases in different mtDNAs.

\section{Frequency spectra}\label{sec:Spectra}

\begin{figure}[h]
\centerline{\includegraphics[scale=0.8]{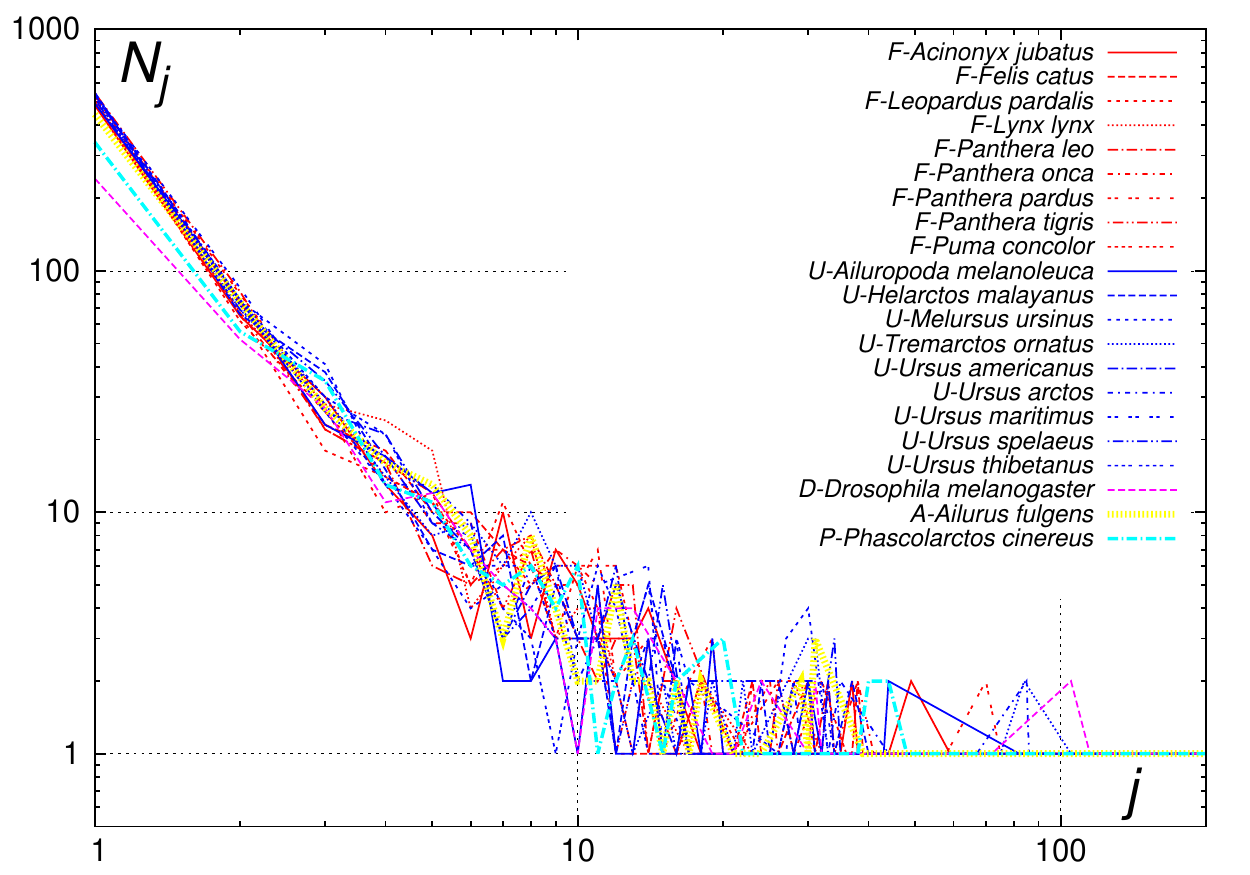}}
\caption{Frequency spectra. The first capital letter before the hyphen serves to distinguish families.
}\label{fig:spectra}
\end{figure}

From a rank--frequency distribution one can obtain the so called frequency spectrum $N_j$, which is the number of items occurring exactly $j$ times \cite{Tuldava:1996,Popescu_etal:2009}. The spectra for nucleotide sequences of species analyzed in the present work are plotted in Fig.~\ref{fig:spectra}.

In the domain of low ranks, frequency spectra of words were shown to satisfy the following model inspired by the Bose-distribution  \cite{Rovenchak&Buk:2011PhysA,Rovenchak&Buk:2011JPS,Rovenchak:2014JQL}:
\be\label{eq:T}
N_j = \frac{1}{z^{-1}X\left(\frac{(j-1)^\gamma}{T}\right)-1}
\qquad\textrm{with\quad $X(t)=e^t$}.
\end{align}
The fugacity analog $z$ is fixed by the number of \textit{hapax legomena} (items occurring only once in a given sample) $N_1$:
\be
z= \frac{N_1}{N_1+1}.
\end{align}
The remaining parameters $\gamma$ and $T$ are obtained by fitting Eq.~(\ref{eq:T}) to the observed data.

As frequency spectra corresponding to word distributions typically have thick tails, a modification of the above model with nonadditive statistics was also developed \cite{Rovenchak:2015CSCS,Rovenchak&Buk:2018}. In particular, one can use $X(t) = \exp_\kappa(t)$, where the $\kappa$-exponential \cite{Kaniadakis:2001,Kaniadakis:2013} is defined as
\be\label{kappa-def}
\exp_\kappa(x) = \left(\sqrt{1+\kappa^2 x^2}+\kappa x\right)^{\frac 1 \kappa}
\end{align}
reducing to the ordinary exponential in the limit of $\kappa\to0$.

\begin{figure}[h]
\centerline{\includegraphics[scale=0.8]{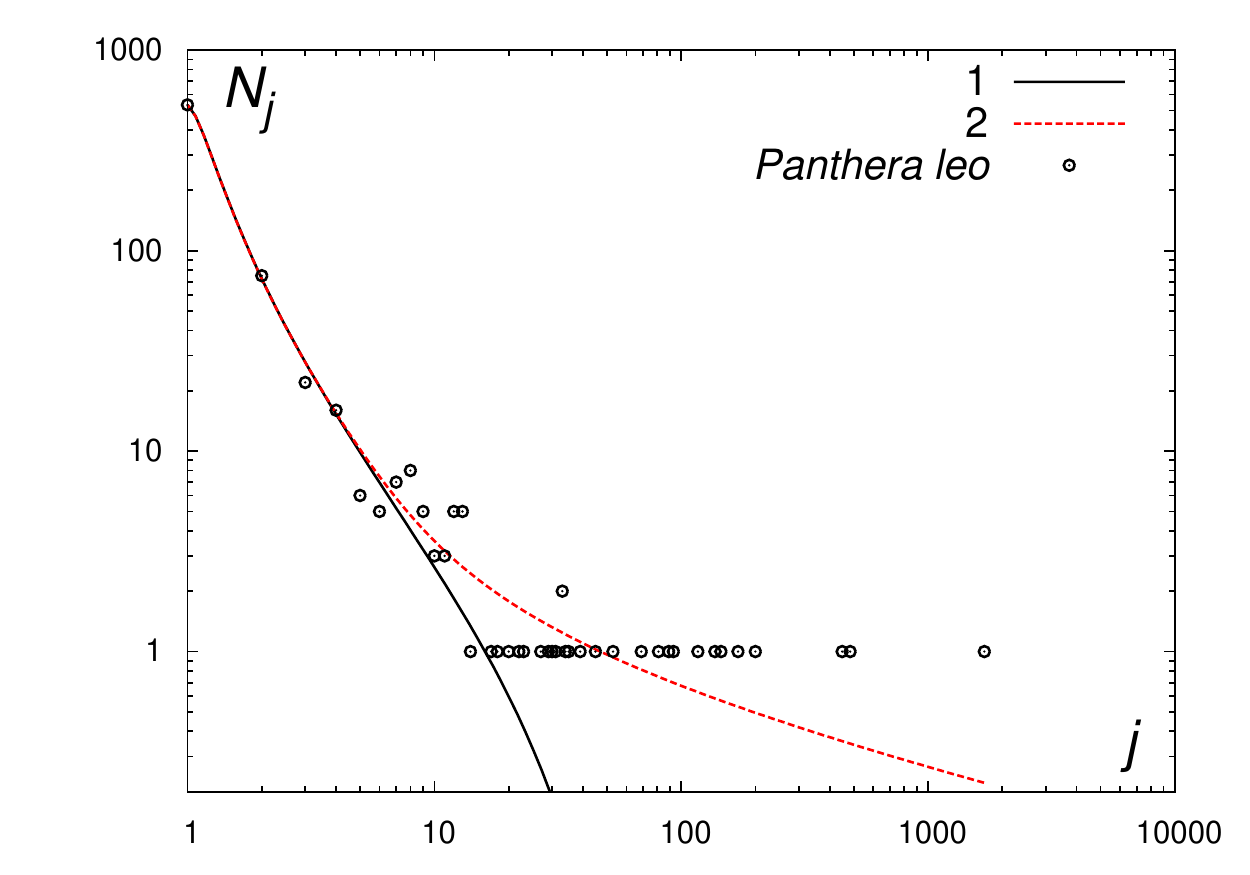}}
\centerline{\includegraphics[scale=0.8]{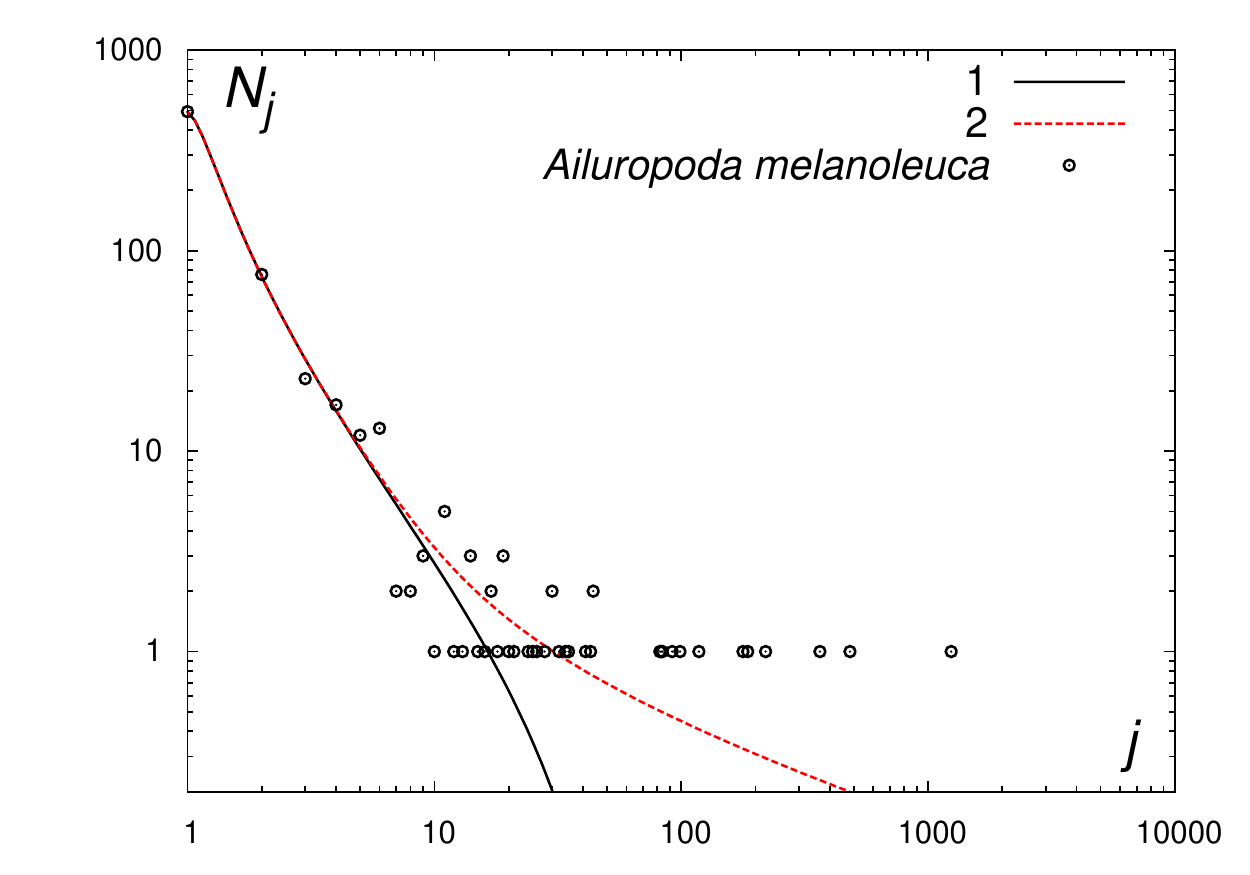}}
\caption{Fitting frequency spectra corresponding to mitochondrial genomes of lion (top panel) and giant panda (bottom panel). Fits with ordinary exponentials are solid lines (1) and fits using $\kappa$-exponentials are dashed lines (2).
}\label{fig:spectrum-fit}
\end{figure}

We have applied this approach to nucleotide sequences. Some results of fitting are demonstrated in Fig.~\ref{fig:spectrum-fit}.

\begin{figure}[h]
\centerline{\includegraphics[scale=0.8]{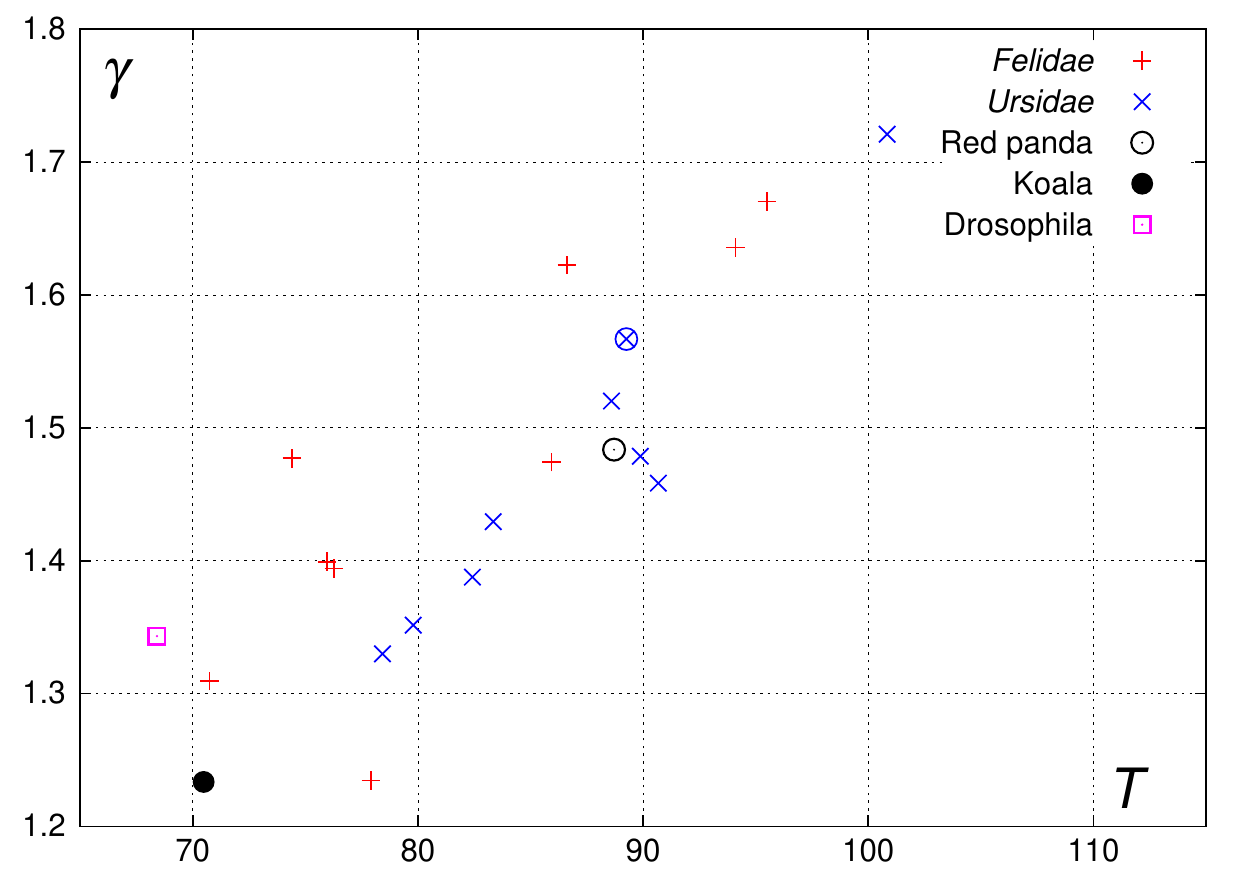}}
\centerline{\includegraphics[scale=0.8]{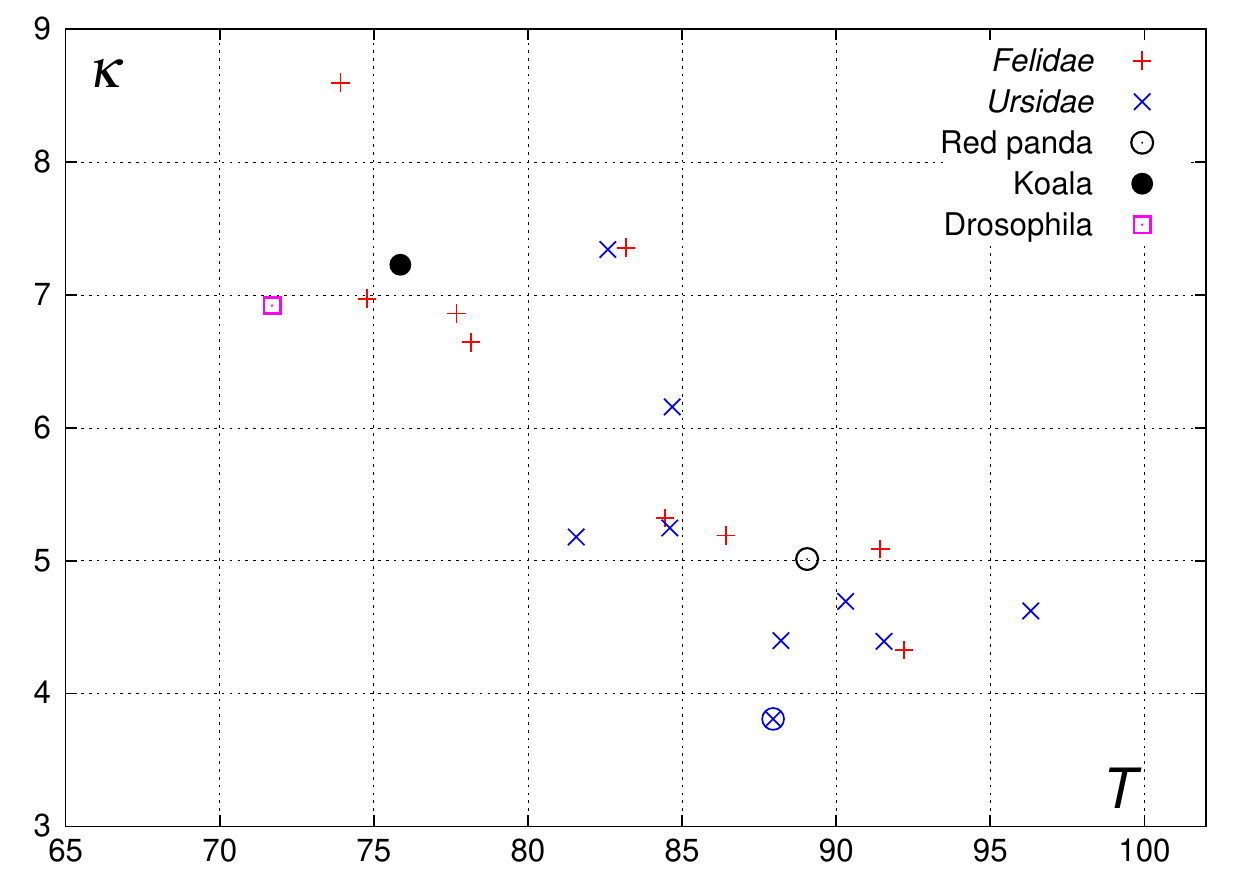}}
\caption{Values of $T$--$\gamma$ (top panel) and $T$--$\varkappa$ (bottom panel) for families and species analyzed in the present work.
}\label{fig:aT-kT}
\end{figure}

Figure~\ref{fig:aT-kT} summarizes the obtained values of parameters for all the species studied in the present work. Eq.~(\ref{eq:T}) with ordinary exponential was fitted to the observed data via two parameters, $\gamma$ and $T$, while the fitting with $\kappa$-exponential was made via $\kappa$ and $T$ at fixed $\gamma=1.5$.

\clearpage

The grouping of different species within families with respect to $\gamma$, $\kappa$, and $T$ parameters is much weaker comparing to the parameters analyzed in Sec.~\ref{sec:Params}, so the former set can be used only as a supplementary discrimination tool.

Still, as we can observe from Fig.~\ref{fig:aT-kT}, cats (\textit{Felidae}) generally have longer tails comparing to bears (\textit{Ursidae}; mean value $\langle\kappa\rangle_{\rm F}=6.3$ versus $\langle\kappa\rangle_{\rm U}=5.1$) but lower ``temperatures'' ($\langle T \rangle_{\rm F}=82$ versus $\langle T \rangle_{\rm U}=87$).

\section{Conclusions}\label{sec:Concl}

An approach was proposed for the analysis of nucleotide sequences in mitochondrial DNA in order to find a set of parameters discriminating taxonomic ranks in the biological classification like families and possibly genera. The approach was tested on two carnivoran families, \textit{Felidae} (cats) and \textit{Ursidae} (bears).

The nucleotide sequences were defined using the linguistic analogy, with the most frequent nucleobase (adenine in all the analyzed cases) as a separating element (a whitespace analog separating nucleotide ``words''). The rank--frequency distributions were compiled, entropy $S$ and mean length $\langle L\rangle$ were calculated. The latter pair of parameters was shown to serve well for the discrimination of cat and bear families. As one of the results, we were able to confirm that \textit{Ailuropoda melanoleuca} (giant panda) is a bear, that \textit{A.~melanoleuca} and \textit{Ailurus fulgens} (red panda) belong to different families, and that \textit{Phascolarctos cinereus} (koala) is not a bear at all.

A linear relation was observed between entropy $S$ and mean length $\langle L\rangle$, which triggered a search for a simplified model describing these parameters. Such a model yielding nearly linear relation for $S$ and $\langle L\rangle$ in the appropriate range of values was found. Further adjustments are required in order to achieve not only a qualitative but also a quantitative agreement with the observed data.

The so called frequency spectra obtained from the rank--frequency distributions were modeled using a nonadditive modification of the Bose-distribution. Such an approach allowed for better description of thick (or long) tails in the spectra. Various parameters describing families and species were obtained. Unlike entropy and mean sequence length, they cannot serve for a decisive separation of animal families. Still, it was found that on average the frequency spectra of \textit{Felidae} (cats) have longer tails than those of the \textit{Urdidae} (bears) family.

In summary, the proposed approaches can be used in studies of mitochondrial genomes as the suggested set of parameters serve to discriminate animal families. Inclusion of other species is planned in future in order to check the applicability of the approaches and to define the ranges of parameters corresponding to families and genera.

\section*{Acknowledgments}
I am grateful to my colleagues, Dr. Volodymyr Pastukhov an Yuri Krynytskyi for inspiring discussions as well as to Dr. Przemko Wa\-li\-szewski for hints regarding genome databases.

This work was partly supported by Project FF-30F (No. 0116U001539) from the Ministry of Education and Science of Ukraine

\bibliographystyle{unsrt}

\end{document}